\documentclass[reprint,amssymb,amsmath,aps,showpacs,10pt,floatfix,longbibliography]{revtex4-1}

\usepackage[pdftex]{graphicx} % Include figure files
\usepackage{epstopdf}
\usepackage{verbatim}
\usepackage{amsmath}
\usepackage{color}
\usepackage{subfigure}
\usepackage{amsbsy}
\usepackage{wasysym}

\begin{document}

\title{Gapped ground state in the zigzag pseudospin-$\frac{1}{2}$ \\ quantum antiferromagnetic chain compound PrTiNbO$_6$}

\author{Yuesheng Li}
\email{yuesheng.man.li@gmail.com}
\author{Sebastian Bachus}
\author{Yoshifumi Tokiwa}
\author{Alexander A. Tsirlin}
\author{Philipp Gegenwart}
\affiliation{Experimental Physics VI, Center for Electronic Correlations and Magnetism, University of Augsburg, 86159 Augsburg, Germany}

\date{\today}

\begin{abstract}
We report a single-crystal study on the magnetism of the rare-earth compound PrTiNbO$_6$ that experimentally realizes the zigzag pseudospin-$\frac{1}{2}$ quantum antiferromagnetic chain model. Random crystal electric field caused by the site mixing between non-magnetic Ti$^{4+}$ and Nb$^{5+}$, results in the non-Kramers ground state quasi-doublet of Pr$^{3+}$ with the effective pseudospin-$\frac{1}{2}$ Ising moment. Despite the antiferromagnetic intersite coupling of about 4 K, no magnetic freezing is detected down to 0.1 K, whilst the system approaches its ground state with almost zero residual spin entropy. At low temperatures, a sizable gap of about 1 K is observed in zero field. We ascribe this gap to off-diagonal anisotropy terms in the pseudospin Hamiltonian, and argue that rare-earth oxides open an interesting venue for studying magnetism of quantum spin chains.
\end{abstract}


\maketitle

\section{Introduction}

One-dimensional (1D) $S$ = $\frac{1}{2}$ quantum spin chains (QSC) have captured extensive interest of both theorists and experimentalists for almost 100 years. Strong quantum fluctuations are integral to such systems, owing to their low-dimensionality, low coordination number, and low spin. Abundant novel phases of matter, including ``quantum spin liquid'' (QSL) phases named Luttinger liquids in the purely 1D case, may occur~\cite{haldane1981luttinger,faddeev1981spin,kolezhuk1998non,balents2010spin,lee2008end,wen2004quantum}. QSC is a basic model, which in special cases can even be solved exactly~\cite{lieb1961two}. It is well known that the $S$ = $\frac{1}{2}$ antiferromagnetical Heisenberg spin chain with only nearest-neighbor (NN) interactions has no long-range order and exhibits a gapless energy spectrum according to the Lieb-Schultz-Mattis Theorem~\cite{lieb1961two}. Magnetic excitations of this model can be represented by pairs of $spinon$s carrying spin-$\frac{1}{2}$ each~\cite{faddeev1981spin}. When Heisenberg interactions decaying as 1/$r^2$ are added beyond NN, a gapless resonating-valence-bond ground state (GS) is expected~\cite{haldane1988exact}. In the presence of the XXZ anisotropy, a quantum-critical and gapless Luttinger liquid has been identified for a large (easy-plane) parameter region~\cite{giamarchi2004quantum,RevModPhys.84.1253,PhysRevB.96.134429}.

On the experimental side, many of the transition-metal compounds manifest different flavors of the QSC physics. Exchange couplings in these compounds can be isotropic, as in Sr$_2$CuO$_3$ (SrCuO$_2$)~\cite{motoyama1996magnetic,azuma1997switching}, Cu$^{2+}$ molecular magnets~\cite{landee2013recent}, CuGeO$_3$~\cite{PhysRevLett.70.3651} and NaTiSi$_2$O$_6$ (LiTiSi$_2$O$_6$)~\cite{isobe2002novel}, or anisotropic, as in several spin-chain Co$^{2+}$ oxides with the strong easy-axis anisotropy~\cite{heid1995magnetic,maartense1977field,hanawa1994anisotropic,he2009cov2o6,mekata1978magnetic,yoshizawa1979neutron,he2005crystal,he2006antiferromagnetic}. Most of such compounds are magnetically ordered at low temperatures~\cite{motoyama1996magnetic,azuma1997switching,landee2013recent,heid1995magnetic,maartense1977field,hanawa1994anisotropic,he2009cov2o6,mekata1978magnetic,yoshizawa1979neutron,he2005crystal,he2006antiferromagnetic}.
The delocalization of the 3$d$ electrons causes non-negligible interchain interactions $J_{\perp}$ that eventually stabilize a long-range-ordered GS~\cite{motoyama1996magnetic,wang2017experimental}.

Recent studies uncovered the diverse and hitherto largely unexplored magnetism of 4$f$ oxides. For example, we reported a triangular QSL candidate YbMgGaO$_4$~\cite{li2015gapless,li2015rare,PhysRevLett.117.097201,PhysRevLett.118.107202,li2017nearest}, where NN magnetic couplings are highly anisotropic in spin space and trigger new physics beyond that of the conventional XXZ model~\cite{zhu2017}. Moreover, localized nature of the 4$f$ electrons renders long-range couplings diminutively small and enhances low-dimensionality. Following the same idea, we seek to arrange rare-earth ions along a zigzag chain, arriving at a quasi-1D antiferromagnet with highly anisotropic intrachain and vanishingly small interchain couplings.

In this paper, we report a comprehensive study of the low-$T$ magnetic properties for the rare-earth zigzag QSC single crystal, PrTiNbO$_6$. Random crystal electric field (CEF) caused by the site mixing between non-magnetic Ti$^{4+}$ and Nb$^{5+}$, results in the non-Kramers GS of Pr$^{3+}$, a quasi-doublet with the effective pseudospin-$\frac{1}{2}$ moment and Ising anisotropy. No conventional spin freezing is detected down to 0.1 K, where residual entropy is close to zero, and the system approaches its GS. A sizable energy gap just above the GS is clearly detected, and its possible origin is discussed.

\section{Experimental details}

\begin{figure}[t]
\begin{center}
\includegraphics[width=8.5cm,angle=0]{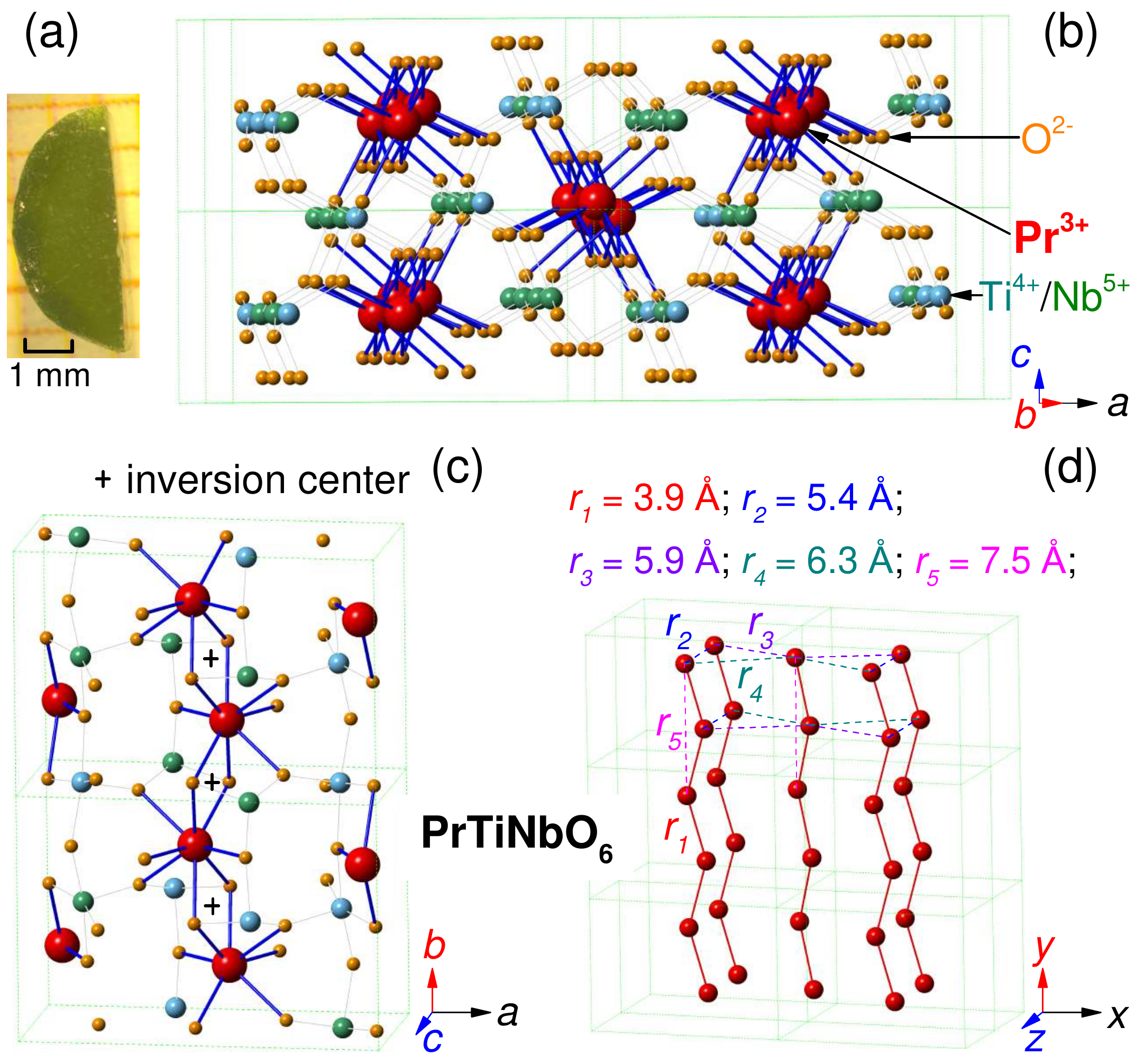}
\caption{(Color online)
(a) A typical as-grown single crystal of PrTiNbO$_6$. (b) View along the Pr$^{3+}$-chain ($b$-axis) of the crystal structure. (c) Zoom-in view of the Pr$^{3+}$-chain in the $ab$-plane. (d) Schematic plot for the Pr$^{3+}$-chains. Ti$^{4+}$, Nb$^{5+}$ and O$^{2-}$ are omitted for clarity, and the coordinate system for the spin components is defined in the inset. The NN Pr$-$Pr distance, $r_1$ = 3.9 {\AA}, is comparable to the shortest Yb$-$Yb distance of 3.4 {\AA}  in YbMgGaO$_4$~\cite{li2015gapless,li2015rare}. The green dashed lines mark the unit cells.}
\label{fig1}
\end{center}
\end{figure}

Polycrystalline samples of Pr$_{x}$La$_{1-x}$TiNbO$_6$ ($x$ = 1, 0.5, 0.2, 0.08, 0.04 and 0) were synthesized by the traditional solid-state method. Large single crystals of PrTiNbO$_6$ [$\sim$ 1 cm, see Fig.~\ref{fig1} (a)] were grown by the optical floating zone technique (see Appendix A). No measurable conductance of the sample was detected at room temperature for PrTiNbO$_6$.

The direct and alternating current (AC) magnetization (1.8 $\leq$ $T$ $\leq$ 400 K, 0 $\leq$ $\mu_0H$ $\leq$ 7 T, and 7.57 $\leq$ $\nu$ $\leq$ 757 Hz) was measured by a magnetic property measurement system (MPMS, Quantum Design) using a 59.77 mg single crystal. The DC magnetization up to 14 T was measured by a vibrating sample magnetometer in a physical property measurement system (PPMS, Quantum Design).

\begin{figure}[t]
\centering
\includegraphics[width=8.5cm,angle=0]{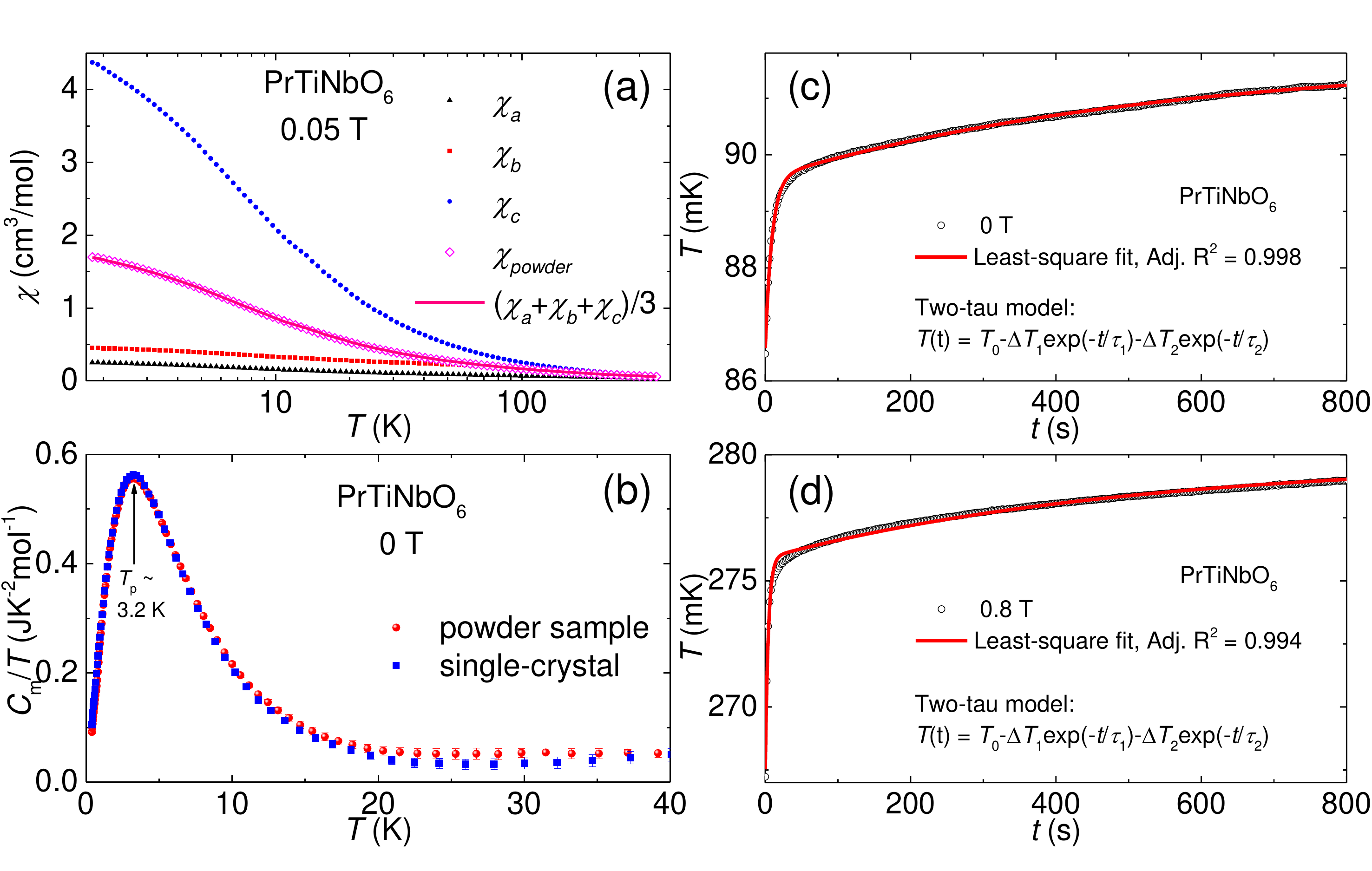}
\caption{(Color online)
(a) Temperature dependence of the magnetic susceptibilities for both single crystal (along the $a$-, $b$-, and $c$-axes) and powder of PrTiNbO$_6$, measured in an applied magnetic field of 0.05 T down to 1.8 K. (b) Low-$T$ magnetic heat capacities measured for both single-crystal and powder samples of PrTiNbO$_6$ under zero field. Relaxation curves of the PrTiNbO6 single crystal measured (c) at 0 T and (d) at 0.8 T applied along the $c$-axis, with the red lines showing the corresponding least-square fits using the two-tau model.}
\label{fig2}
\end{figure}

The heat capacity (0.36 $\leq$ $T$ $\leq$ 400 K and 0 $\leq$ $\mu_0H$ $\leq$ 10 T) was measured using cold-pressed powders ($\sim$ 10 mg), as well as a small single crystal of PrTiNbO$_6$ (5.37 mg) at 0 T, in a PPMS. N-grease was used to facilitate thermal contact between the sample and the puck below 210 K, while H-grease was used above 200 K. The sample coupling was better than 95\%. The contributions of the grease and puck under different external fields were measured independently and subtracted from the data. The low temperatures down to 0.36 K were achieved by a $^3$He refrigerator. The low-$T$ heat capacity of the PrTiNbO$_6$ single crystal down to 0.09 K was measured in a home-built setup installed in a $^3$He -$^4$He dilution refrigerator at external magnetic fields up to 0.8 T applied along the $c$-axis. No significant sample dependence of the thermodynamic properties was observed for both powders and single crystals of PrTiNbO$_6$ [see Fig.~\ref{fig2} (a) and (b)].

The setup for the milli-K heat capacity measurements is shown in the inset of Fig.~\ref{fig6} (b)~\cite{tokiwa2011}. The single-crystal sample used in the measurement was cut into a small thin plate (5.37 mg). Its largest face was well polished and contacted to the platform using grease. Thermal link to the bath was achieved by a bronze wire of appropriate diameter and length to produce quasi-adiabatic conditions. Thermal relaxation was analyzed using the two-$\tau$ model similar to Ref.~\onlinecite{design2004ppms},
\begin{multline}
P_h(t)=C_{\rm pl}\frac{dT_{\rm pl}}{dt}+K_g[T_{\rm pl}(t)-T_{\rm sa}(t)],\\
K_g[T_{\rm pl}(t)-T_{\rm sa}(t)]=C_{\rm sa}\frac{dT_{\rm sa}}{dt}+K_w[T_{\rm sa}(t)-T_{\rm ba}],
\label{eq1}
\end{multline}
where $C_{\rm pl}$ and $C_{\rm sa}$ are the heat capacities of the platform and sample respectively, $K_g$ is the thermal conductance between the two, $K_w$ is the thermal conductance of the thermal link, $T_{\rm pl}$, $T_{\rm sa}$, and $T_{\rm ba}$ are the temperatures of the platform, the sample, and the thermal bath, respectively. And $P_h(t)$ is the power of the heater. The solution to the model takes the form,
\begin{equation}
T_{sa}(t)=T_0-\Delta T_1\, e^{-t/\tau_1}-\Delta T_2\,e^{-t/\tau_2}.
\label{eq2}
\end{equation}
The experimental data can be fitted using Eq.~(\ref{eq2}) with five independent variables, $T_0$, $\Delta T_1$, $\Delta T_1$, $\tau_1$, and $\tau_2$ [see Fig.~\ref{fig2} (c) and (d)]. Below 2 K, the background heat capacity ($\sim$ $C_{\rm pl}$) was negligible, $C_{\rm pl}\sim 0.01$\,$\mu$J\,K$^{-1}$ $<0.03 C_{\rm sa}$. Thus, we can obtain the temperature and heat capacity as $T=T_0-\Delta T_1/2-\Delta T_2/2$ and $C_{\rm sa}$ = $P(\tau_1+\tau_2)/(\Delta T_1+\Delta T_2)$, respectively, according to Eq.~(\ref{eq1}). Here $P$ is the power increase of the heater.

At low temperatures, the nuclear contribution becomes dominant (see below), and the experimental data slightly deviate from the two-$\tau$ model at $\sim$ 30 s [see Fig.~\ref{fig2} (c)]. At higher applied fields, the deviations occur even at higher temperatures [see Fig.~\ref{fig2} (d) for example]. This deviation may be caused by the thermal decoupling between the phonon (lattice) and electronic/nuclear subsystems~\cite{smith2005origin}. To avoid errors in the resulting heat capacity, we chose to exclude the heat capacity data with the adj. $R^2$~\footnote{See https://www.originlab.com/doc/Origin-Help/Interpret-Regression-Result for adj. $R^2$} smaller than 0.998 [see Fig.~\ref{fig2} (d) for example]. Therefore, our zero-field data can extend to the lowest temperature of 0.09 K, whereas the data obtained in non-zero magnetic fields terminate at higher temperatures. This is because the applied magnetic field increases the nuclear gap due to Zeeman effect, and the nuclear contribution to the heat capacity shifts to higher temperatures. It also serves as a probe of the local magnetization, see Sec.~\ref{sec:nuclear} for further analysis.

\section{Results and discussion}

\subsection{Crystal structure}

PrTiNbO$_6$ belongs to the family of RTiNbO$_6$ oxides (R = La, Ce, Pr, Nd and Sm) with the space group $Pnma$ (see Fig.~\ref{fig1}). This family has been extensively studied due to the promising applications in ceramic microwave resonators and miniature solid-state laser systems~\cite{sebastian2001preparation,surendran2003microwave,qi1997modified,qi1996potential,qi1996optical}. However, its low-$T$ magnetism was not reported to date.

The superexchange coupling between the NN Pr$^{3+}$ ions (Wyckoff position, $4c$) is mediated by two parallel anions O$^{2-}$(2) ($8d$) along the zigzag chain (see Fig.~\ref{fig1} and Appendix A). Large spatial separation between the chains renders the magnetism quasi-1D [see Fig.~\ref{fig1} (b)]. Assuming \mbox{dipole-dipole} nature of long-range couplings, we estimate the interchain interaction $J_{\perp}$ $\leq$ $\mu_0\mu_B^2g_c^2$/(16$\pi$$r_2^3$) $\sim$ 0.02 K, and the next-nearest-neighbor intrachain interaction $J_{2}$ $\leq$ $\mu_0\mu_B^2g_c^2$/(16$\pi$$r_5^3$) $\sim$ 0.006 K, whereas the NN intrachain coupling is about 4 K, as we show below.

The site mixing between the magnetic and nonmagnetic ions is unlikely due to the large chemical differences between Pr$^{3+}$ with the ionic radius $r_{\rm Pr^{3+}}$ = 0.99 {\AA} and Ti$^{4+}$/Nb$^{5+}$ ions with $r_{\rm Ti^{4+}/Nb^{5+}}$ = 0.61/0.64 {\AA}~\cite{li2015gapless}. On the other hand, Ti$^{4+}$ and Nb$^{5+}$ are statistically distributed in the structure. Our x-ray diffraction data confirm the statistical mixing and the absence of any superstructure reflections (see Fig.~\ref{fig8}). No Pr$^{3+}$ defects were detected either.

In the PrTiNbO$_6$ structure, inversion centers located halfway between the Pr$^{3+}$ ions [see Fig.~\ref{fig1} (c)] exclude the antisymmetric Dzyaloshinsky-Moriya interactions~\cite{moriya1960new,li2015gapless}. This renders PrTiNbO$_6$ a promising system for studying 1D anisotropic magnetism in the zigzag-chain geometry.

\begin{figure}[t]
\centering
\includegraphics[width=8.5cm,angle=0]{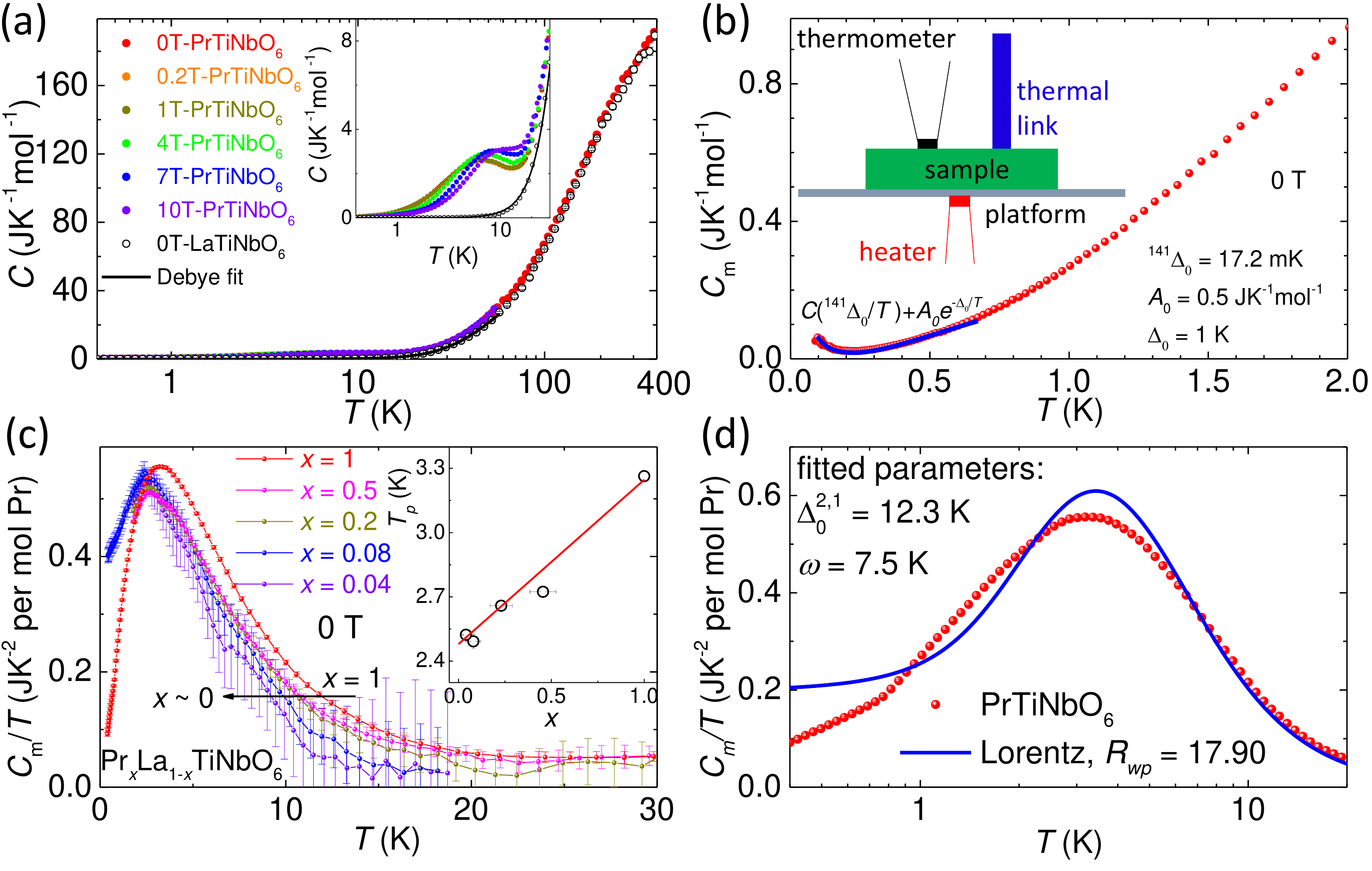}
\caption{(Color online)
(a) Heat capacities measured on the powders of PrTiNbO$_6$ and LaTiNbO$_6$ down to 0.4 K under different magnetic fields. The black curve presents a Debye fit to the heat capacity of LaTiNbO$_6$ below 50 K with the fitted Debye temperature $\sim$ 260 K. The inset shows a zoom-in view of the low-$T$ data. (b) Low-$T$ magnetic heat capacities of PrTiNbO$_6$ measured at 0 T down to 0.09 K. The blue curve shows the fit below 0.7 K with the fitted parameters shown in the figure. The inset shows the schematic view of the platform used for the milli-K heat capacity measurement. (c) Magnetic heat capacity measured on the Pr$_{x}$La$_{1-x}$TiNbO$_6$ ($x$ = 1, 0.5, 0.2, 0.08, and 0.04) powders. The inset shows the $x$ dependence of the peak position of the hump ($T_p$), with the red line representing a linear fit. (d) Magnetic heat capacity measured on PrTiNbO$_6$ with the blue line showing the least-$R_{wp}$ fit with the Lorentzian distribution of $E_2$-$E_1$.}
\label{fig6}
\end{figure}

\subsection{Crystal-field randomness and the ground-state quasi-doublet}

An isolated Pr$^{3+}$ ion with the electronic configuration 4$f^2$ forms the ninefold-degenerate GS, $\mid\!m_J\rangle$ ($m_J$ = $\pm$4, $\pm$3, $\pm$2, $\pm$1, 0), with the spin angular momentum $s$ = 1, orbital angular momentum $L$ = 5, and total angular momentum $J$ = 4, according to the Hund's rules. In PrTiNbO$_6$, the $C_{1h}$ point-group symmetry of the Pr$^{3+}$ site splits this GS into nine singlets.

\begin{figure}[t]
\begin{center}
\includegraphics[width=8.5cm,angle=0]{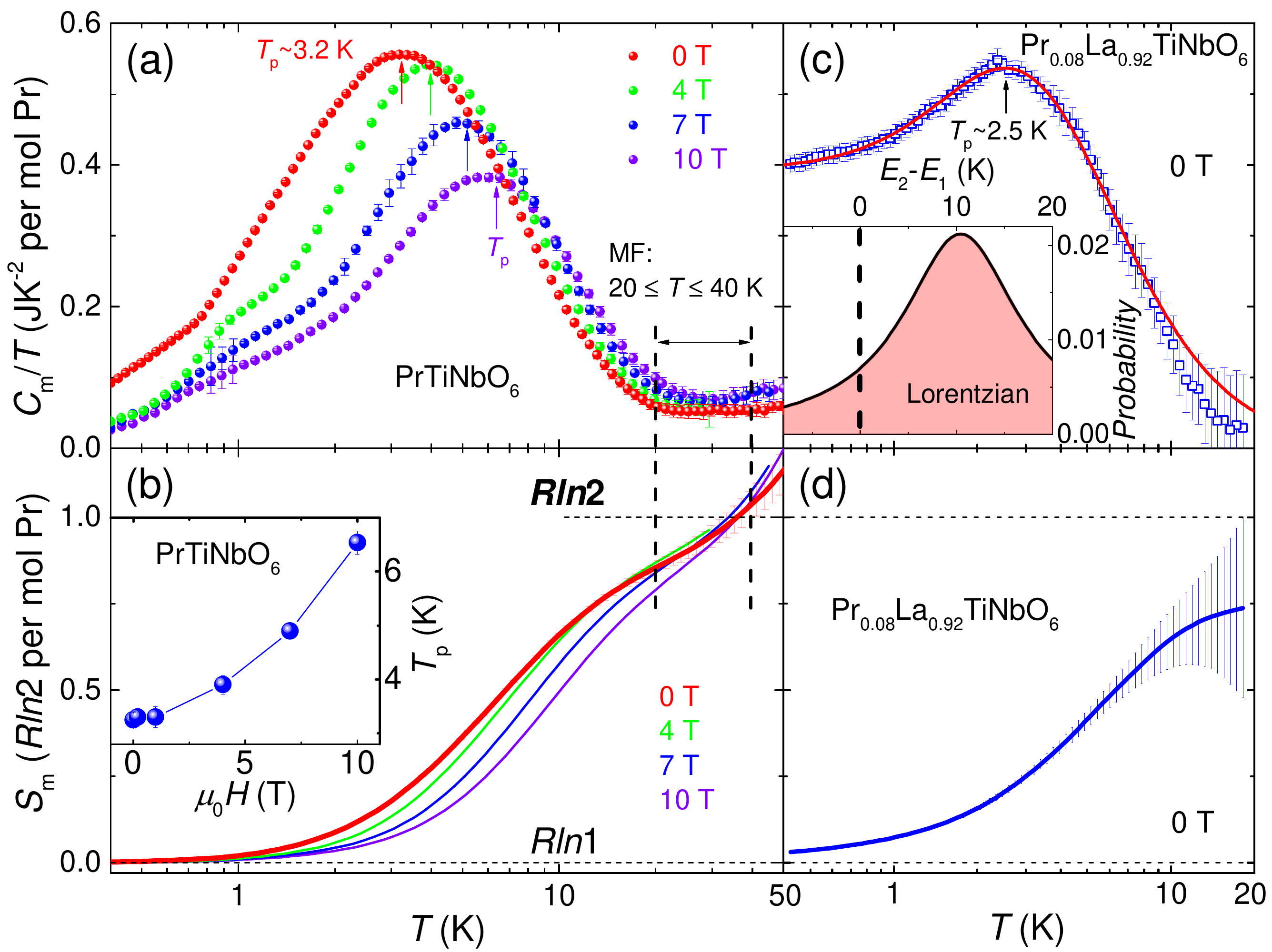}
\caption{(Color online)
(a) Magnetic heat capacity of the PrTiNbO$_6$ powder measured under external magnetic fields. The hump center temperatures ($T_p$) are marked. (b) Magnetic entropy of PrTiNbO$_6$. The inset shows the field dependence of $T_p$. (c) Magnetic heat capacity of the Pr$_{0.08}$La$_{0.92}$TiNbO$_6$ powder, with the red line showing the Lorentz fit. The inset represents the resulting Lorentzian distribution of $E_2-E_1$ with the fitted parameters, $\Delta_0^{2,1}$ = 10.5 K (center) and $\omega$ = 15 K (width). (d) Magnetic entropy of Pr$_{0.08}$La$_{0.92}$TiNbO$_6$.}
\label{fig3}
\end{center}
\end{figure}

The magnetic heat capacity ($C_m$) of PrTiNbO$_6$ can be determined very accurately by subtracting the heat capacity of the non-magnetic LaTiNbO$_6$  as phonon contribution [see Fig.~\ref{fig6} (a)]. At 0.4 K, the magnetic heat capacity of PrTiNbO$_6$ is extremely small, $C_m$ = 0.006$R\ln2$, suggesting that the ground-state regime has been reached. At higher temperatures, a broad hump of $C_m$/$T$ centered at $T_p$ $\sim$ 3.2 K is observed. By integrating $C_m$/$T$ [Fig.~\ref{fig3} (a)] over $T$, we obtain the magnetic entropy [Fig.~\ref{fig3} (b)] that sharply increases in the vicinity of $T_p$. The curve flattens out around 40 K, where the value of $R\ln2$ is reached. At higher temperatures, the entropy increases again due to the population of higher-lying CEF levels.

The nature of the magnetic entropy can be two-fold. The CEF excitations between the two lowest-lying CEF singlets (single-ion magnetism) on one hand and the intersite magnetic interactions (cooperative magnetism) on the other hand, both contribute to the heat capacity and produce the total entropy of $R\ln2$. To distinguish between these effects, we measured heat capacity of the magnetically diluted samples Pr$_{x}$La$_{1-x}$TiNbO$_6$. At $x$ = 0.04, where intersite interactions should be fully suppressed, the entire hump is gradually shifted to a lower temperature of $\sim$ 2.5 K [see Fig.~\ref{fig6} (c)]. However, we were unable to describe this hump assuming excitations between the two lowest CEF singlets, $\mid\!E_1\rangle$ and $\mid\!E_2\rangle$.

In a two-level system,
\begin{multline}
C_m^{\rm CEF}=R\int p(E_2-E_1-\Delta_0^{2,1})(\frac{E_2-E_1}{T})^2\\
\frac{e^{\frac{|E_2-E_1|}{T}}}{(1+e^{\frac{|E_2-E_1|}{T}})^2}d(E_2-E_1),
\label{eq3}
\end{multline}
where $p(E_2-E_1-\Delta_0^{2,1})$ = $\delta(E_2-E_1-\Delta_0^{2,1})$ is the spectral weight distribution function for the two-level model, $\Delta_0^{2,1}$ is the energy gap center, and $(E_2-E_1)$ is the integration variable. The fit quality is assessed using
\begin{equation}
R_{wp}=\sqrt{\frac{1}{N}\sum_{j=1}^{N}\left(\frac{Y_j^{\rm obs}-Y_j^{\rm cal}}{\sigma_j^{\rm obs}}\right)^2},
\label{eq4}
\end{equation}
resulting in very high values $R_{wp}$ = 21 and 6.4 for $x=0.08$ and 0.04, respectively. Here, $Y_j^{\rm obs}$ is the observed value, $\sigma_j^{\rm obs}$ is its standard deviation and $N$ is the number of the data points, whereas $Y_j^{\rm cal}$ is the calculated value. Indeed, the finite value of $C_m$/$T$ $\sim$ 0.4 JK$^{-2}$ per mol Pr at 0.4 K [see Fig.~\ref{fig3} (c)] would not be possible in a two-level system, because such a system is necessarily gapped.

In YbMgGaO$_4$, complete site mixing between Mg$^{2+}$ and Ga$^{3+}$ was reported to generate random electric fields on the Yb$^{3+}$ ions and produce a broad distribution of the CEF levels. This distribution has the Lorentzian width of about 5 meV, as probed by inelastic neutron scattering~\cite{PhysRevLett.118.107202}. A similar effect should take place in the RTiNbO$_6$ family, where the complete site mixing between Ti$^{4+}$ and Nb$^{5+}$ ~\cite{sebastian2001preparation,surendran2003microwave,qi1997modified,qi1996potential,qi1996optical} generates random electric fields on the rare-earth sites. Taking this possibility into account, we modeled the heat capacity of the diluted samples assuming a Lorentzian distribution of the CEF excitation energy, $E_2$-$E_1$,
\begin{equation}
p(E_2-E_1-\Delta_0^{2,1})=\frac{2}{\pi}\frac{\omega}{4(E_2-E_1-\Delta_0^{2,1})^2+\omega^2}.
\label{eq5}
\end{equation}
A perfect fit of the data for Pr$_{0.08}$La$_{0.92}$TiNbO$_6$ ($R_{wp}=0.38$) yields the maximum of the distribution at $\Delta_0^{2,1}$ = 10.5 K and the width $\omega$ = 15 K [see Fig.~\ref{fig3} (c)]. With $\omega$ $>$ $\Delta_0^{2,1}$, an accidental degeneracy of $|E_1\rangle$ and $|E_2\rangle$ occurs, and even at 0.4 K the heat capacity remains finite. A similar fit of the data for Pr$_{0.04}$La$_{0.96}$TiNbO$_6$ ($R_{wp}=0.52$) produced nearly the same parameters, $\Delta_0^{2,1}$ = 10.3 K and $\omega$ = 17 K (see Appendix B), thus confirming that the mixing of Pr$^{3+}$ and La$^{3+}$ has minor effect on the CEF levels. In contrast to Ti$^{4+}$ and Nb$^{5+}$, the La$^{3+}$ and Pr$^{3+}$ ions have the same point charge along with similar ionic radius. Therefore, they should not cause any significant changes in the local electric field environment of Pr$^{3+}$. Note also that in YbMgGaO$_4$ charge misbalance due to the mixing of differently charged ions, Mg$^{2+}$ and Ga$^{3+}$, is the major cause of local structural distortions and ensuing CEF randomness~\cite{PhysRevLett.118.107202}.

The picture emerging so far is that of a rather unexpected but mundane single-ion CEF physics. A striking observation is that the undiluted sample, PrTiNbO$_6$, features a much lower heat capacity at 0.4 K compared to the diluted sample with $x$ = 0.08. This leads us to conclude that the two lowest CEF states mix into a quasi-doublet, $|\sigma_{\pm}\rangle=\frac{1}{\sqrt{2}}(|E_1\rangle\pm|E_2\rangle)$, and render Pr$^{3+}$ a \mbox{pseudospin-$\frac12$} magnetic ion. Interactions between such ions open a gap and reduce the low-$T$ heat capacity with respect to its single-ion value, compare panels (a) and (c) of Fig.~\ref{fig3}. On the other hand, we were unable to fit the data for the undiluted sample using the single-ion CEF model, either without or with the CEF randomness [see Fig.~\ref{fig6} (d)].

\begin{figure}[t]
\begin{center}
\includegraphics[width=8.5cm,angle=0]{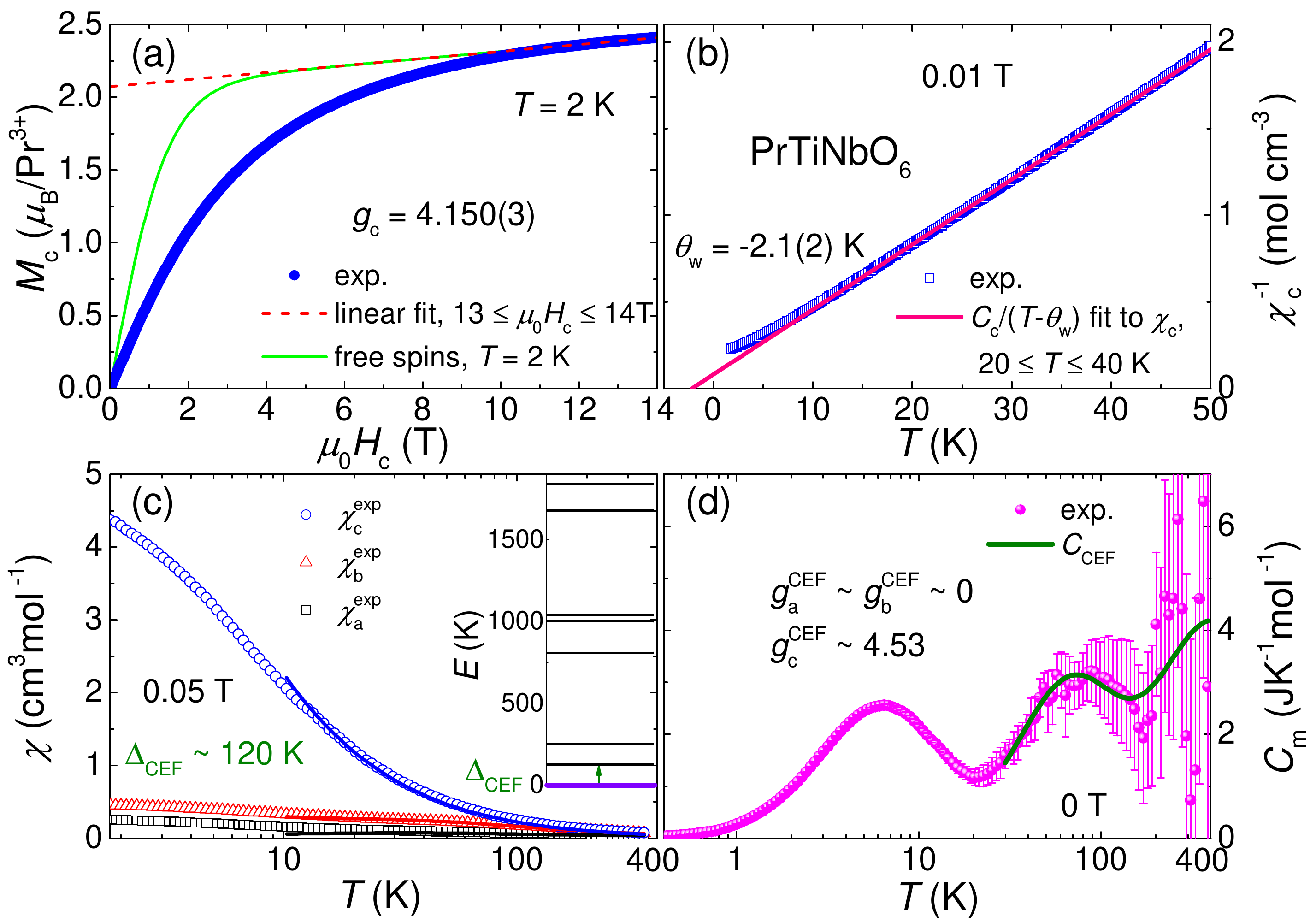}
\caption{(Color online)
(a) Magnetization of the PrTiNbO$_6$ single crystal measured along the $c$-axis at 2 K. The red dashed line shows the linear fit to the data from 13 to 14 T, and the green line represents the calculation for the Pr$^{3+}$ spins without intersite couplings. (b) Inverse magnetic susceptibility measured in the field of 0.01 T along the $c$-axis. The red line shows the Curie-Weiss fit from 20 to 40 K. Combined CEF fit to both (c) susceptibilities (along the $a$, $b$ and $c$-axes) and (d) heat capacity above 30 K. The inset of (c) shows the resulting CEF energy-levels with the GS quasi-doublet (violet line) and 7 excited singlets (black lines).}
\label{fig4}
\end{center}
\end{figure}

\begin{figure}[t]
\centering
\includegraphics[width=8.5cm,angle=0]{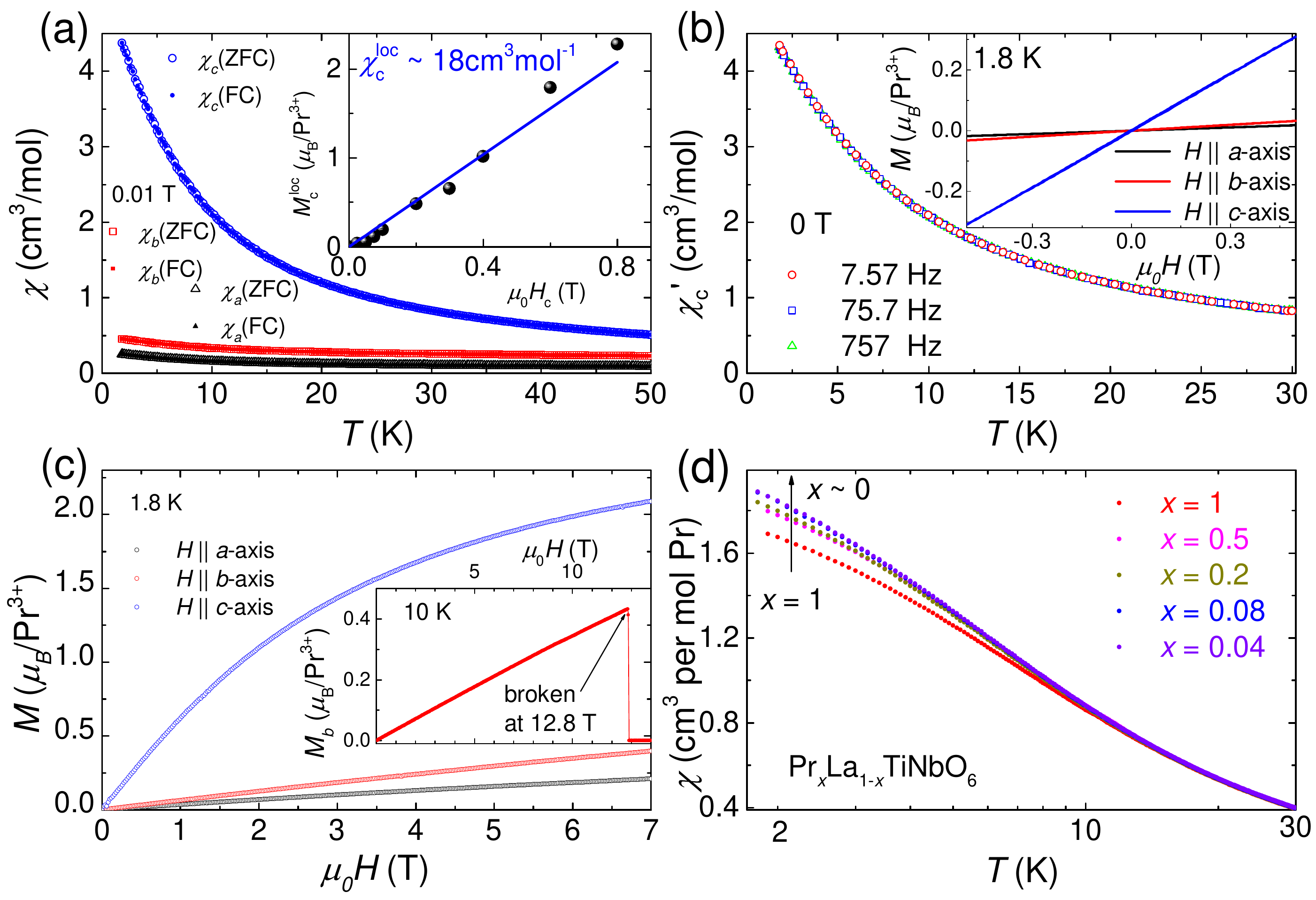}
\caption{(Color online)
(a) Magnetic susceptibilities measured under zero-field cooling (ZFC) and field cooling (FC) down to 1.8 K for the single crystal of PrTiNbO$_6$ along the $a$-, $b$-, and $c$-axes. The inset represents the local magnetization extracted from the low-$T$ heat capacity with the blue line showing a linear fit. (b) AC susceptibilities measured for the single crystal of PrTiNbO$_6$ along the $c$-axis under zero field down to 1.8 K. The inset shows the complete magnetization loops measured at 1.8 K along the $a$-, $b$-, and $c$-axes. (c) Field dependence of the magnetization measured at 1.8 K along the $a$-, $b$-, and $c$-axes. The inset shows the magnetization along the $b$-axis at 10 K up to 12.8 T. (d) Magnetic susceptibility of the Pr$_{x}$La$_{1-x}$TiNbO$_6$ powders ($x$ = 1, 0.5, 0.2, 0.08, and 0.04).}
\label{fig7}
\end{figure}

Further fingerprints of the intersite interactions between such magnetic entities are seen in the magnetization data. At 2 K and above $\sim$ 10 T, the magnetization shows full polarization only along the $c$-axis with a pseudospin-$\frac{1}{2}$ $g$-factor, $g_c$ $\sim$ 4.15, and a very small Van Vleck susceptibility $\chi_c^{\rm vv}$ $\sim$ 0.16 cm$^3$/mol [see Fig.~\ref{fig4} (a)]. The magnetization along the $a$- or $b$-axis, $M_a$ or $M_b$, is proportional to the applied magnetic field up to 7 T [see Fig.~\ref{fig7} (c)]. It was not possible to measure $M_a$ and $M_b$ in much higher applied fields, as the strong field broke the sample due to the strong magnetic anisotropy [inset of Fig.~\ref{fig7} (c)]. Because $C_m$/$T$ reaches a minimum between 20 and 40 K [see Fig.~\ref{fig3} (a)], both the intersite spin-spin correlations and CEF excitations have marginal effect in this temperature range, and it can be used for the Curie-Weiss fitting. We thus obtained $\theta_w$ = -2.1(2) K, and the effective moment of $\mu_{\rm eff}$ = 2.37(2)$\mu_{B}$ $\sim$ $\frac{g_c}{2}\mu_{B}$ [see Fig.~\ref{fig4} (b)]. Therefore, the intersite coupling is clearly antiferromagnetic, $J_{zz}$ = -2$\theta_w$ $\sim$ 4 K, as further confirmed by modeling the magnetization at 2 K, where strong deviations from the magnetization process of an uncoupled Pr$^{3+}$ entity are observed [see Fig.~\ref{fig4} (a)].

Below $T_p$, the susceptibility $\chi_c$ is lower than predicted by the Curie-Weiss law [see Fig.~\ref{fig4} (b)], suggesting that intersite antiferromagnetic correlations set in. The value of $T_{p}$ increases with the applied magnetic field [see Fig.~\ref{fig3} (b)], similar to YbMgGaO$_4$~\cite{li2015gapless}. Furthermore, the AC susceptibility shows no frequency-dependence down to 1.8 K, thus excluding spin freezing below $T_{p}$ [see Fig.~\ref{fig7}].

\subsection{CEF Hamiltonian}

In order to probe higher-lying CEF states, we use the generic CEF Hamiltonian with the $C_{1h}$ symmetry~\cite{feldmann1975crystal,PhysRevB.88.214419,bauer2009magnetism} and exclude CEF randomness. The data are fitted above 30 K, where effects of mixing between the two lowest states and of intersite correlations are excluded. We performed a combined single-ion CEF fit (see Appendix C) to $\chi_a$, $\chi_b$, $\chi_c$ [Fig.~\ref{fig4} (c)], and $C_m$ [Fig.~\ref{fig4} (d)] taking into account the pseudospin-$\frac{1}{2}$ $g$-factors ($g_a=g_b=0$, $g_c\sim4.15$) from the magnetization data, and $\langle E_2$-$E_1\rangle$ $\sim$ 10 K [see Fig.~\ref{fig3} (c)]. The lowest CEF level is found at $\Delta_{\rm CEF}$ $\sim$ 120 K above the GS quasi-doublet.

\subsection{Effective pseudospin-$\frac{1}{2}$ Hamiltonian}

At low temperatures, the formation of the pseudospin-$\frac{1}{2}$ quasi-doublet with the strong Ising anisotropy~\cite{nekvasil1990effective} resembles non-Kramers GS doublets in the 3D pyrochlore compounds Pr$_2$TM$_2$O$_7$ (TM = Sn, Zr, Hf, and Ir), where Ising anisotropy had been reported too~\cite{matsuhira2002low,matsuhira2004low,PhysRevLett.101.227204,PhysRevB.88.104421,PhysRevLett.114.017602}.
It is thus logical to use a similar NN pseudospin-$\frac{1}{2}$ Hamiltonian for interactions along the zigzag chain. The projection of the dominant NN superexchange interaction through the virtual $fpf$ hopping process ~\cite{PhysRevLett.105.047201,PhysRevB.83.094411,onoda2011effective} onto the subspace of the GS CEF quasi-doublet ($|\sigma_{\pm}\rangle$) results in,
\begin{multline}
\mathcal{H}=\sum_{k=1}^{\infty}[J_{zz}S_k^zS_{k+1}^z+J_{\pm}(S_k^+S_{k+1}^-+S_k^-S_{k+1}^+)\\
+J_{\pm\pm}(S_k^+S_{k+1}^++S_k^-S_{k+1}^-)\\
+i^{2k+1}J_{\pm\pm}'(S_k^+S_{k+1}^+-S_k^-S_{k+1}^-)],
\label{eq6}
\end{multline}
with the time-reversal invariant quadrupole moment $S_k^{\pm}$ $\equiv$ $S_k^{x}\pm iS_k^{y}$ in the non-Kramers case~\cite{PhysRevLett.105.047201,PhysRevB.83.094411,onoda2011effective,PhysRevB.86.104412,PhysRevB.94.205107}. The contribution of dipole-dipole interactions is negligible, $\mu_0\mu_B^2g_c^2$/(16$\pi$$r_1^3)$ $\sim$ 0.01$J_{zz}$.

\begin{figure}[t]
\centering
\includegraphics[width=8cm,angle=0]{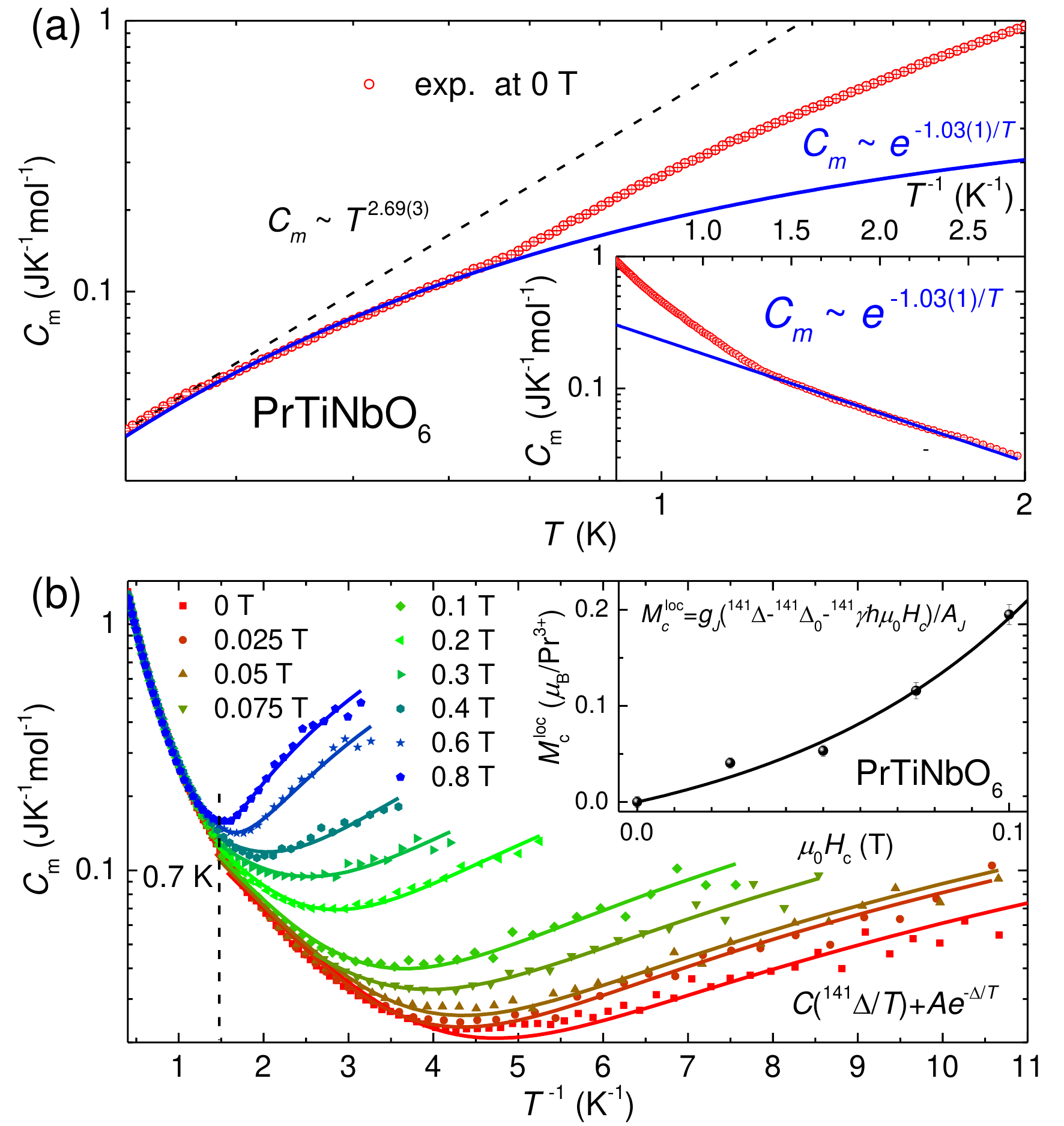}
\caption{(Color online)
(a) Low-$T$ magnetic heat capacities of PrTiNbO$_6$ measured at 0 T using the PPMS $^3$He option. The black dashed line shows the power-law dependence, while the blue solid line represents the exponential fit below 0.7 K. The inset shows the corresponding $C_m$ vs. $T^{-1}$ plot. The nuclear contribution is negligible above 0.3 K at 0 T. (b) Magnetic heat capacities of the PrTiNbO$_6$ single crystal measured in magnetic fields applied along the $c$-axis using the home-built setup in a $^3$He -$^4$He dilution refrigerator, with the colored lines showing the corresponding fits below $\sim$ 0.7 K. The inset shows field dependence of the local magnetization below 0.1 T with the black line representing an exponential fit.}
\label{fig5}
\end{figure}

\subsection{Low-energy spin excitation}
\label{sec:nuclear}
We now turn to the spin gap of PrTiNbO$_6$. The exponential behavior of the zero-field magnetic heat capacity is observed at 0.36 $\leq T\leq$ 0.7 K. By fitting the data with $C_m$ $\propto$ $\exp(-\Delta_0/T$), we arrive at $\Delta_0$ = 1.03(1) K $\sim$ 0.2$J_{zz}$ [see Fig.~\ref{fig5} (a)]. Alternatively, one could fit the data assuming a power-law behavior ($\sim$ $T^\gamma$), but such a fit extends only up to 0.4 K with a large exponent, $\gamma$ = 2.7, thus confirming the gapped behavior [see Fig.~\ref{fig5} (a)]. Moreover, the very small magnetic heat capacity, $C_m< 0.004 R\ln2$ below 0.4 K [see Fig.~\ref{fig5} (b)], ensures the formation of a spin gap. Residual spin entropy, $S_m\leq 0.001R\ln2$, indicates that the GS behavior is approached at $\sim$ 0.3 K. The single-ion CEF excitations show a gapless behavior in the highly diluted samples (see Fig.~\ref{fig3}). Thus the gapped behavior of PrTiNbO$_6$ is due to cooperative magnetism, namely, the spin-spin interactions along the chain.

Neither easy-plane nor easy-axis spin-$\frac{1}{2}$ XXZ chain models~\cite{PhysRevB.57.11429,RevModPhys.84.1253} can account for the gapped and disordered GS observed in PrTiNbO$_6$, indicating the importance of the $J'_{\pm\pm}$ and $J_{\pm\pm}$ terms in Eq.~\eqref{eq6}. It is worth noting that the $J'_{\pm\pm}$ term breaks the translational symmetry, following zigzag geometry of the spin chain. In this case, an energy gap may open, according to the Haldane conjecture~\cite{bulaevskii1963theory,PhysRevB.7.3166,PhysRevB.46.3486,totsuka1997magnetization,PhysRevB.58.9156,paul2017ground,haldane1983continuum,haldane1983nonlinear,affleck1989quantum}.

Finally, let us discuss the zero-field heat capacity below 0.3 K. In the absence of the nuclear contribution, magnetic heat capacity should rapidly decrease to zero with decreasing temperature, $\sim$ exp(-$\Delta_0$/$T$). However, experimentally it starts increasing at $\sim$ 0.2 K [see Fig.~\ref{fig5} (b)]. In external fields applied along the $c$-axis, this tail gradually shifts towards higher temperatures with an effective gyromagnetic ratio,
\begin{equation}
\frac{\gamma_c^{\rm eff}(\mu_0H_c)}{2\pi} = \frac{k_B[^{141}\Delta(\mu_0H_c)-^{141}\!\Delta_0]}{h\mu_0H_c},
\label{eq7}
\end{equation}
where $^{141}\Delta_0$ and $^{141}\Delta(\mu_0H_c)$ are the energy gaps of the Schottky tails at applied magnetic fields of 0 T and $\mu_0H_c$, respectively. $\gamma_c^{\rm eff}{\rm (0.8\,T)}/(2\pi)$ $\sim$ 3.9 GHz/T is one order of magnitude smaller than the gyromagnetic ratio of free electrons and excludes the effect of free (defect) Pr$^{3+}$ electronic spins in general. On the other hand, $\gamma_c^{\rm eff}/(2\pi)$ is two orders of magnitude larger than the gyromagnetic ratio of free $^{141}$Pr nuclear spins, $^{141}\gamma/(2\pi)$ $\equiv$ 13 MHz/T, and comparable to the hyperfine coupling constant of Pr$^{3+}$, $A_J/h$ $\equiv$ 1093 MHz~\cite{Bak1992nmr,PhysRevLett.76.1936}. Therefore, we expect that the low-$T$ tail in the heat capacity originates from the $^{141}$Pr nuclear spins, which are hyperfine-coupled to the local magnetization of Pr$^{3+}$ electronic spins, $M_c^{\rm loc}$. The similar tails were also reported as zero-field nuclear contribution in other QSL candidates~\cite{PhysRevLett.117.267202,NC2011Gapless}.

To avoid the possible artifact of the fitting procedure, we fit the low-$T$ magnetic heat capacity with $C_m=C(^{141}\Delta/T)+A\exp(-\Delta/T)$ below $\sim$ 0.7 K [see Fig.~\ref{fig5} (b)], where the first term is the heat capacity of the $^{141}$Pr nuclear spins expressed by a two-level model [see Eq.~\eqref{eq3}]. The increase in $^{141}\Delta$ under the magnetic field measures the low-$T$ $M_c^{\rm loc}$ by~\cite{PhysRevLett.76.1936}
\begin{equation}
M_c^{\rm loc}=\frac{g_J}{A_J}(^{141}\Delta-^{141}\Delta_0-^{141}\gamma\hbar\mu_0H_c),
\label{eq8}
\end{equation}
where $^{141}\Delta_0$ = 17.2(2) mK is the nuclear gap at 0 T [see Fig.~\ref{fig6} (b)], and $g_J$ = 4/5 is the Land\'{e} $g$-factor. Below $\mu_0H_c$ = 0.1 T $\sim$ $0.1J_{zz}/(g_c\mu_B)$, $M_c^{\rm loc}$ shows an exponential-like field dependence [see Fig.~\ref{fig5} (b)], possibly confirming the aforementioned gapped behavior. Above 0.1 T, $M_c^{\rm loc}$ depends on the field almost linearly, and at 0.8 T [$\sim$ 0.6$J_{zz}/(g_c\mu_B)$] nearly reaches the full polarization~\footnote{It is not advisable to measure the heat capacity of the PrTiNbO$_6$ single crystal under larger external magnetic fields, as the magnetic anisotropy is too large, $M_c$ $\gg$ $M_a$ or $M_b$, and the torque may be strong enough to destroy the device.}, indicating an enhanced spin susceptibility at low-$T$ and above 0.1 T, $\chi_c^{loc}$ = 18(1) cm$^3$mol$^{-1}$ $\sim$ 4$\chi_c$(1.8 K) [see Fig.~\ref{fig7} (a)].

\section{Conclusions and Outlook}

We reported the formation of a non-trivial magnetic state in the spin-chain compound PrTiNbO$_6$. CEF randomness caused by the site mixing between Ti$^{4+}$ and Nb$^{5+}$, results in the non-Kramers GS quasi-doublet of Pr$^{3+}$ with Ising anisotropy. The zigzag chains of the pseudospin-$\frac{1}{2}$ Pr$^{3+}$ ions feature antiferromagnetic couplings $J_{zz}\simeq$ 4 K, and reveal a sizable spin gap, $\sim$ 0.2$J_{zz}$. The gap opening may be related to off-diagonal anisotropy terms that alternate along the chain according to its zigzag geometry. Future studies of spin dynamics and correlations of PrTiNbO$_6$ are warranted and made possible through the availability of high-quality single crystals. More generally, we establish quasi-1D rare-earth oxides as a playground for exploring magnetism of quantum spin chains.

\acknowledgements

We thank Haijun Liao, Yuanpai Zhou, and Gang Chen for helpful discussion, Prof. Winzer for advice on the single-crystal growth, and Sebastian Esser for his technical help in G\"ottingen. This work was supported by the German Science Foundation through TRR-80 and the German Federal Ministry for Education and Research through the Sofja Kovalevskaya Award of Alexander von Humboldt Foundation.

\section*{Appendix}

\appendix

\section{Sample synthesis and characterization}

\begin{figure}[h]
\centering
\includegraphics[width=8.5cm,angle=0]{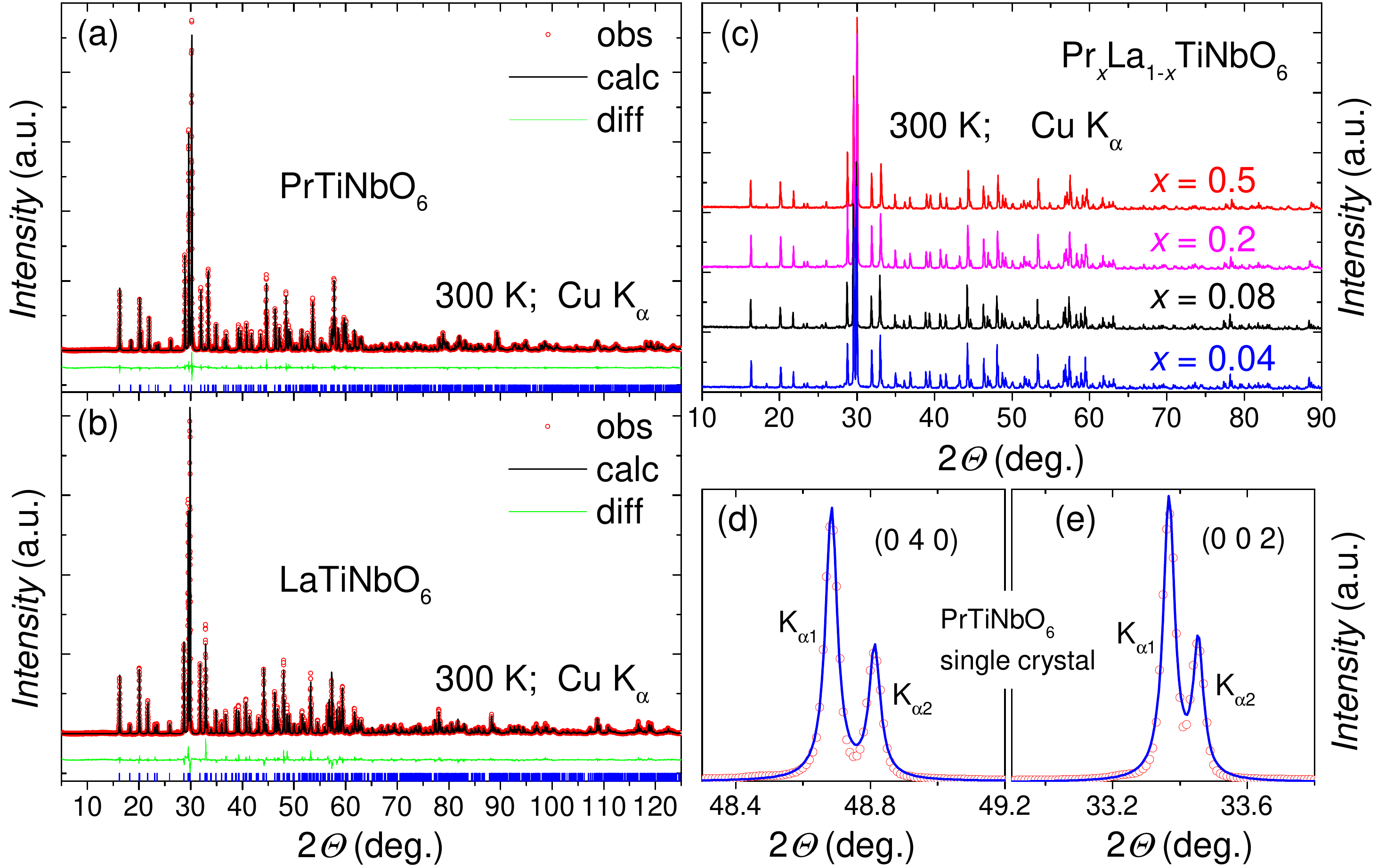}
\caption{(Color online)
X-ray powder diffraction and Rietveld refinement patterns (a) for PrTiNbO$_6$ and (b) for LaTiNbO$_6$. (c) X-ray powder diffraction patterns for Pr$_{x}$La$_{1-x}$TiNbO$_6$ ($x$ = 0.5, 0.2, 0.08, and 0.04). Bragg reflections of the PrTiNbO$_6$ single crystal (d) on the $ac$ - plane, (0 4 0), with a FWHM $\sim$ 0.05$^o$, and (e) on the $ab$ - plane, (0 0 2), with a FWHM $\sim$ 0.05$^o$.}
\label{fig8}
\end{figure}

\begin{figure}[t]
\centering
\includegraphics[width=8.5cm,angle=0]{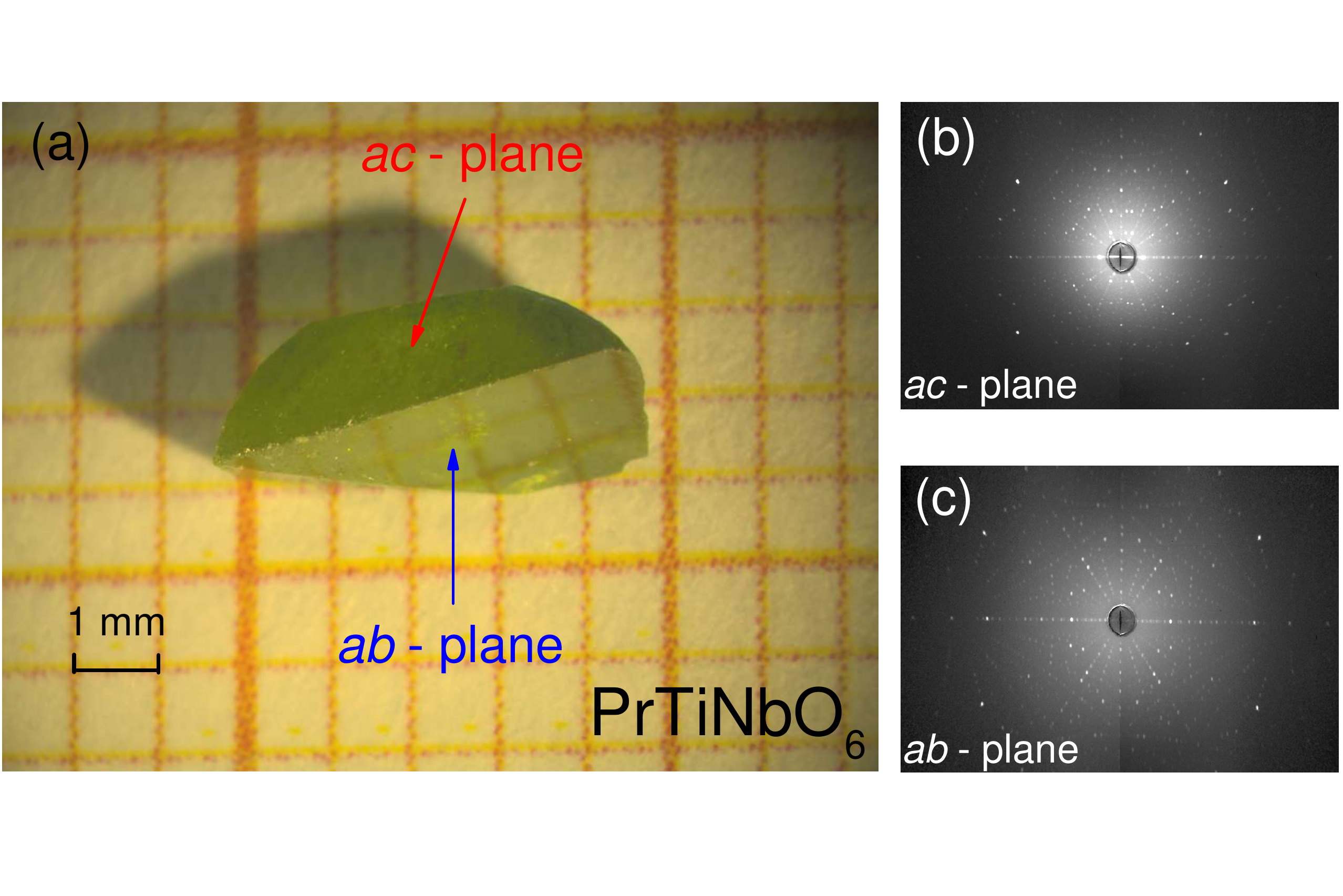}
\caption{(Color online)
(a) A single crystal of PrTiNbO$_6$ grown by the floating zone technique. The cut planes, $ac$-plane and $ab$-plane, are marked. Laue x-ray diffraction patterns (b) on the $ac$-plane and (c) on the $ab$-plane.}
\label{fig9}
\end{figure}

The polycrystalline samples of Pr$_{x}$La$_{1-x}$TiNbO$_6$ ($x$ = 1, 0.5, 0.2, 0.08, 0.04, and 0) were synthesized by a traditional solid-state method. The stoichiometric mixtures of Pr$_6$O$_{11}$, La$_2$O$_{3}$, TiO$_2$, and Nb$_2$O$_5$ were heated in air up to 1350 $^{\circ}$C for 2 days, with an intermediate grinding. An annealing down to 1080 $^{\circ}$C was performed for Pr$_{x}$La$_{1-x}$TiNbO$_6$ ($x$ = 0.2, 0.08, 0.04, and 0) in order to obtain the right phase. From $x$ = 1 to $x$ = 0, the color of the Pr$_{x}$La$_{1-x}$TiNbO$_6$ powders gradually changes from green to white. Phase purity was confirmed by x-ray diffraction measurements [see Fig.~\ref{fig8} (a), (b), and (c)].

Crystal structures of both PrTiNbO$_6$ and LaTiNbO$_6$ were refined based on the x-ray diffraction data using the Rietveld method implemented in the General Structure Analysis System (GSAS) program [see Fig.~\ref{fig8} (a) and (b)]~\cite{larson1994gsas}. The refined structures are shown in Table~\ref{table1} for both PrTiNbO$_6$ and LaTiNbO$_6$, which are consistent with the previously reported results~\cite{sebastian2001preparation,surendran2003microwave,qi1997modified,qi1996potential,qi1996optical}. No signatures of preferred orientation were observed [see Fig.~\ref{fig8} (a) and (b)].

\begin{table}[t]
\caption{Rietveld-refined crystal structures of PrTiNbO$_6$ and LaTiNbO$_6$ at $\sim$ 300 K.}\label{table1}
\begin{center}
\begin{tabular}{ l | l || l | l }
    \hline
    \hline
    \multicolumn{2}{l||}{ model: $Pnma$ } & PrTiNbO$_6$ & LaTiNbO$_6$ \\ \hline
    lattice & \emph{a} & 10.96779(7) & 10.94373(9) \\
    parameters & \emph{b} & 7.52340(5) & 7.58623(6) \\
    {\AA} & \emph{c} & 5.38146(4) & 5.45202(5) \\ \hline
    Pr$^{3+}$ or La$^{3+}$ & \emph{x} & 0.04242(4) & 0.04302(6) \\
    $4c$ & \emph{y} & 0.25 & 0.25 \\
      & \emph{z} & 0.04091(9) & 0.04138(14) \\
      & $U_{iso}$($\times$100) & 0.56(2) & 0.75(2) \\ \hline
    Ti$^{4+}$ / Nb$^{5+}$ & \emph{frac} & 0.5 & 0.5 \\
    $8d$ & \emph{x} & 0.35550(6) & 0.35663(8) \\
      & \emph{y} & 0.00634(8) & 0.00545(12) \\
      & \emph{z} & 0.03667(14) & 0.03609(20) \\
      & $U_{iso}$($\times$100) & 0.45(2) & 0.63(3) \\ \hline
    O1$^{2-}$ & \emph{x} & 0.2874(3) & 0.2983(4) \\
    $8d$ & \emph{y} & 0.5573(4) & 0.5464(5) \\
      & \emph{z} & 0.3706(6) & 0.3822(8) \\
      & $U_{iso}$($\times$100) & 0.27(6) & 0.62(8) \\ \hline
    O2$^{2-}$ & \emph{x} & 0.0276(3) & 0.0313(4) \\
    $8d$ & \emph{y} & 0.5374(4) & 0.5317(5) \\
      & \emph{z} & 0.2691(5) & 0.2669(7) \\
      & $U_{iso}$($\times$100) & 0.27(6) & 0.62(8) \\ \hline
    O3$^{2-}$ & \emph{x} & 0.1471(3) & 0.1587(5) \\
    $4c$ & \emph{y} & 0.25 & 0.25 \\
      & \emph{z} & 0.4539(9) & 0.4776(12) \\
      & $U_{iso}$($\times$100) & 0.27(6) & 0.62(8)  \\ \hline
    O4$^{2-}$ & \emph{x} & 0.3769(4) & 0.3777(5) \\
    $4c$ & \emph{y} & 0.25 & 0.25 \\
      & \emph{z} & 0.1552(8) & 0.1352(10) \\
      & $U_{iso}$($\times$100) & 0.27(6) & 0.62(8)  \\
    \hline
    \hline
\end{tabular}
\end{center}
\end{table}

Large-size and high-quality green transparent single crystals of PrTiNbO$_6$ [$\sim$ 1 cm, see Fig.~\ref{fig9} (a)] were grown by a high-temperature optical floating zone furnace (FZ-T-10000-H-\uppercase\expandafter{\romannumeral6}-VPM-PC, Crystal Systems Corp.), using 44.4\% of the full power of the four lamps (the full power is 1.5 kW for each lamp). The technical details had been reported in our previous growths for YbMgGaO$_4$ and LuMgGaO$_4$ single crystals~\cite{li2015rare}. The single crystal was oriented by the Laue x-ray diffraction, and was cut consequently by a line cutter along the crystallographic $ac$-plane ($ab$-plane). The properly cut planes (the $ac$ and $ab$-planes) were polished and cross-checked by both conventional [Fig.~\ref{fig8} (d) and (e)] and Laue x-ray diffraction [Fig.~\ref{fig9} (b) and (c)]. The high quality of the crystal was confirmed by the narrow reflection peaks, with full width at half maximum (FWHM), 2$\Delta\Theta$ $\sim$ 0.05$^o$ [see Fig.~\ref{fig8} (d) and (e)].

\section{CEF randomness in Pr$_{x}$La$_{1-x}$TiNbO$_6$}

\begin{figure}[t]
\centering
\includegraphics[width=8.5cm,angle=0]{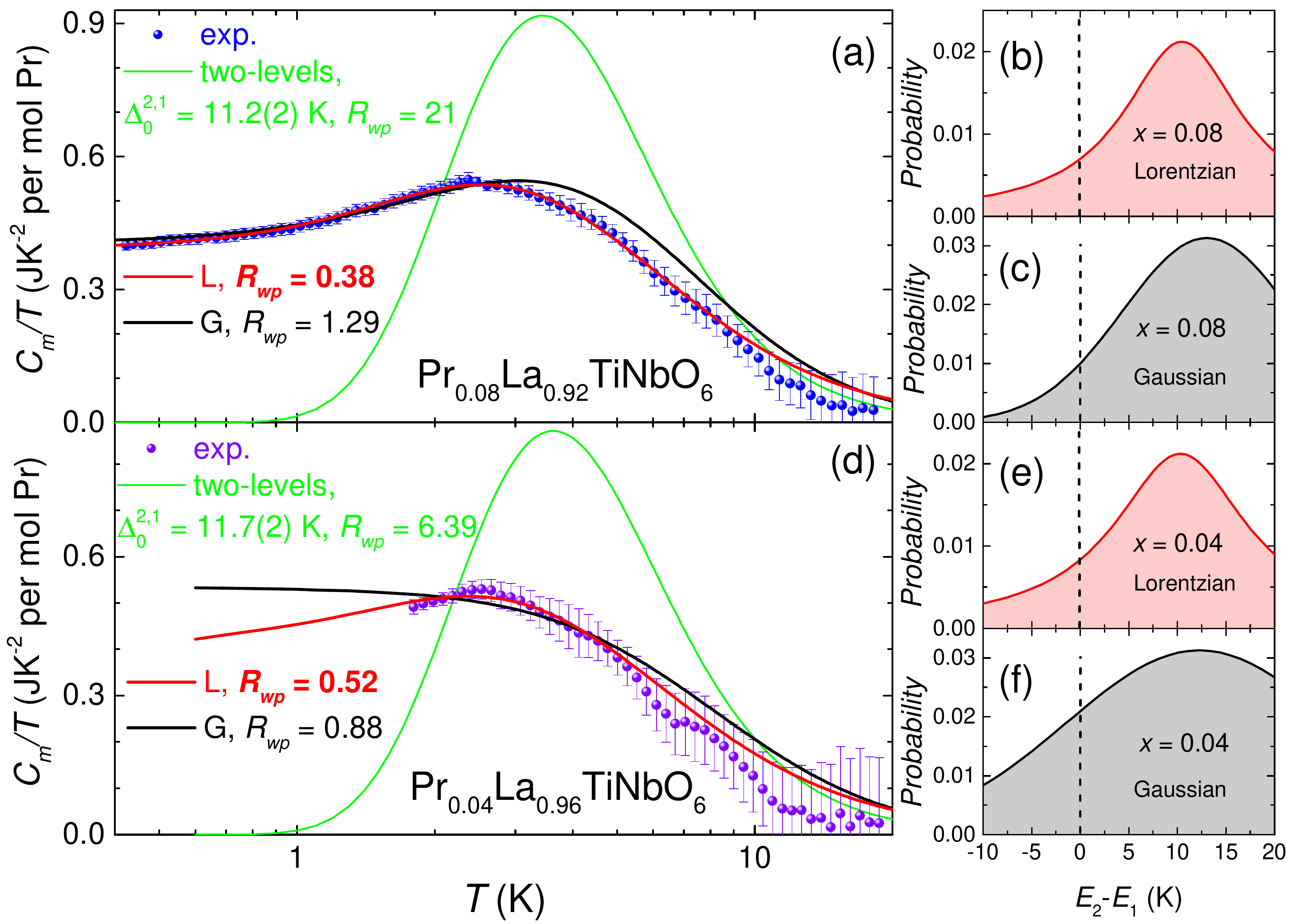}
\caption{(Color online)
Magnetic heat capacity measured on (a) Pr$_{0.08}$La$_{0.92}$TiNbO$_6$ and (d) Pr$_{0.04}$La$_{0.96}$TiNbO$_6$. The green lines show the least-$R_{wp}$ fit by the two-level model without any broadening of $E_2$-$E_1$. The red and black lines represent the least-$R_{wp}$ fits with the Lorentzian (L) and Gaussian (G) broadening of $E_2$-$E_1$, respectively. Resulting Lorentzian distributions of the spectral weight with the fitted parameters, (b) $\Delta_0^{2,1}$ = 10.5 K and $\omega$ = 15 K for Pr$_{0.08}$La$_{0.92}$TiNbO$_6$, and (e) $\Delta_0^{2,1}$ = 10.3 K and $\omega$ = 17 K for Pr$_{0.04}$La$_{0.96}$TiNbO$_6$. Resulting Gaussian distributions of the spectral weight with the fitted parameters, (c) $\Delta_0^{2,1}$ = 13.0 K and $\omega$ = 20 K for Pr$_{0.08}$La$_{0.92}$TiNbO$_6$, and (f) $\Delta_0^{2,1}$ = 12.3 K and $\omega$ = 32 K for Pr$_{0.04}$La$_{0.96}$TiNbO$_6$. No internal CEF gap of the ground-state quasi-doublet is observed.}
\label{fig10}
\end{figure}

The fitting procedure using Lorentzian distribution of $E_2$-$E_1$ is described in the main text. We also tried a similar fit with the Gaussian distribution. Similarly, no internal gap between $\mid\!E_1\rangle$ and $\mid\!E_2\rangle$ is observed [see Fig.~\ref{fig10} (c) and (f)] from the corresponding least-$R_{wp}$ fits. Furthermore, we also tried to use the Lorentzian model, which allowed a good description of the diluted samples [see Fig.~\ref{fig10} (a) and (d)], to fit the magnetic heat capacity of PrTiNbO$_6$. However, this single-ion CEF fit is really poor even if the parameters of the Lorentzian distribution are varied [see Fig.~\ref{fig6} (d)]. This further confirms that the magnetic heat capacity of PrTiNbO$_6$ can not be understood in terms of the single-ion CEF physics, and intersite correlations between the Pr$^{3+}$ spins are important.

\section{Details of CEF analysis}

\begin{table}[t]
\caption{CEF parameters, $B_{n}^{m}$, obtained from the combined fit. All values are in meV.}\label{table2}
\begin{center}
\begin{tabular}{ l | l | l | l | l | l | l | l | l }
    \hline
    \hline
$B_{2}^{0}$ & $B_{2}^{2}$ & $B_{4}^{0}$ & $B_{4}^{2}$ & $B_{4}^{4}$ & $B_{6}^{0}$ & $B_{6}^{2}$ & $B_{6}^{4}$ & $B_{6}^{6}$ \\
    \hline
 1.15 & 0.93 & 0.054 & 0.055 & -0.091 & -0.00053 & -0.0016 & -0.0041 & -0.0019 \\
    \hline
    \hline
\multicolumn{3}{l|}{ } & $B_{2}^{-2}$ & $B_{4}^{-2}$ & $B_{4}^{-4}$ & $B_{6}^{-2}$ & $B_{6}^{-4}$ & $B_{6}^{-6}$ \\
\multicolumn{3}{l|}{ } & $\times10^{8}$ & $\times10^{8}$ & $\times10^{9}$ & $\times10^{9}$ & $\times10^{10}$ & $\times10^{10}$ \\
    \hline
\multicolumn{3}{l|}{ } & 8.3 & 9.5 & 7.9 & -1.4 & -6.8 & -5.4 \\
    \hline
    \hline
\end{tabular}
\end{center}
\end{table}

The generic CEF Hamiltonian with the $C_{1h}$ point group  symmetry at the Pr$^{3+}$ sites is given by~\cite{feldmann1975crystal,PhysRevB.88.214419,bauer2009magnetism},
\begin{multline}
\mathcal{H}_{\rm CEF}=B_{2}^{0}O_{2}^{0}+B_{2}^{2}O_{2}^{2}+B_{2}^{-2}O_{2}^{-2}+B_{4}^{0}O_{4}^{0}+B_{4}^{2}O_{4}^{2}\\
+B_{4}^{-2}O_{4}^{-2}+B_{4}^{4}O_{4}^{4}+B_{4}^{-4}O_{4}^{-4}+B_{6}^{0}O_{6}^{0}+B_{6}^{2}O_{6}^{2}+B_{6}^{-2}O_{6}^{-2}\\
+B_{6}^{4}O_{6}^{4}+B_{6}^{-4}O_{6}^{-4}+B_{6}^{6}O_{6}^{6}+B_{6}^{-6}O_{6}^{-6},
\label{eq9}
\end{multline}
where $B_{n}^{m}$ ($m,n$ are integers and $|m|\leq n$) are CEF parameters, and $O_{n}^{m}$ are the Stevens operators. The eigenvalues and eigenvectors of Eq.~(\ref{eq9}) are given by $E_{i}$ and $\mid\!E_i\rangle$ ($i$ = 1 $-$ 9), respectively. Under a small external magnetic field of $H$ along the $a$-, $b$- or $c$-axis, the CEF Hamiltonian can be expressed by,
\begin{equation}
\mathcal{H}_{\rm CEF}^\alpha=\mathcal{H}_{\rm CEF}-\mu_0\mu_Bg_JHJ_\alpha,
\label{eq10}
\end{equation}
with $\alpha$ = $x$, $y$, and $z$ respectively. The eigenvalues and eigenvectors of Eq.~(\ref{eq10}) are given by $E_{i}^{\alpha}$ and $\mid\!i,\alpha\rangle$, respectively. The single-ion DC magnetic susceptibility can be calculated by,
\begin{equation}
\chi_\alpha^{\rm CEF}=\frac{\mu_Bg_JN_A\sum_{i=1}^{9} exp(-\frac{E_{i}^{\alpha}}{k_BT})\langle i,\alpha|J_\alpha|i,\alpha\rangle}{H\sum_{i=1}^{9}exp(-\frac{E_{i}^{\alpha}}{k_BT})},
\label{eq11}
\end{equation}
and the single-ion magnetic heat capacity under 0 T can be calculated by,
\begin{equation}
C_m^{\rm CEF}=\frac{N_A}{k_{B}T^{2}}\frac{\partial^2\ln[\sum_{i=1}^{9} \exp(-\frac{E_i}{k_{B}T})]}{\partial(\frac{1}{k_{B}T})^2}.
\label{eq12}
\end{equation}

Through a combined fit using Eq.~(\ref{eq9})$-$(\ref{eq12}) to the magnetic susceptibilities and heat capacity measured above 30 K (including the measured pseudospin-$\frac{1}{2}$ $g$-factors and $\langle E_2-E_1\rangle$, see the main text) by minimizing the $R_{wp}$ [see Eq.~(\ref{eq4})],
all of the CEF parameters, $B_{n}^{m}$, can be determined experimentally (see Table~\ref{table2}). Thus, all of the nine eigenvalues (the relative values) and eigenvectors of Eq.~(\ref{eq9}) can be obtained (see Table~\ref{table3}).

\begin{table}[t]
\caption{Fitted CEF energy levels and the corresponding CEF states under 0 T.}\label{table3}
\begin{center}
\begin{tabular}{ l }
    \hline
    \hline
 $E_1$ = 0 K \\
 $|E_1\rangle$ = 0.66$|3\rangle$-0.66$|$-3$\rangle$-0.24$|1\rangle$+0.24$|$-1$\rangle$ \\ \hline
 $E_2$ = 9.4 K \\
 $|E_2\rangle$ = 0.71$|3\rangle$+0.71$|$-3$\rangle$-0.04$|1\rangle$-0.04$|$-1$\rangle$ \\ \hline
 $E_3$ = 125 K \\
 $|E_3\rangle$ = -0.07$|4\rangle$-0.07$|$-4$\rangle$+0.70$|2\rangle$+0.70$|$-2$\rangle$-0.08$|0\rangle$ \\ \hline
 $E_4$ = 250 K \\
 $|E_4\rangle$ = 0.11$|4\rangle$-0.11$|$-4$\rangle$-0.70$|2\rangle$+0.70$|$-2$\rangle$ \\ \hline
 $E_5$ = 810 K \\
 $|E_5\rangle$ = 0.04$|3\rangle$+0.04$|$-3$\rangle$+0.71$|1\rangle$+0.71$|$-1$\rangle$ \\ \hline
 $E_6$ = 1000 K \\
 $|E_6\rangle$ = 0.32$|4\rangle$+0.32$|$-4$\rangle$+0.09$|2\rangle$+0.09$|$-2$\rangle$+0.88$|0\rangle$ \\ \hline
 $E_7$ = 1040 K \\
 $|E_7\rangle$ = 0.24$|3\rangle$-0.24$|$-3$\rangle$+0.66$|1\rangle$-0.66$|$-1$\rangle$ \\ \hline
 $E_8$ = 1680 K \\
 $|E_8\rangle$ = 0.70$|4\rangle$-0.70$|$-4$\rangle$+0.11$|2\rangle$-0.11$|$-2$\rangle$ \\ \hline
 $E_9$ = 1840 K \\
 $|E_9\rangle$ = 0.62$|4\rangle$+0.62$|$-4$\rangle$+0.04$|2\rangle$+0.04$|$-2$\rangle$-0.46$|0\rangle$ \\
    \hline
    \hline
\end{tabular}
\end{center}
\end{table}

The CEF randomness caused by the site mixing between Ti$^{4+}$ and Nb$^{5+}$~\cite{PhysRevLett.118.107202} mixes the two lowest CEF eigenstates, $|E_1\rangle$ and $|E_2\rangle$, because the averaged energy difference, $\langle E_2-E_1\rangle$ $\sim$ 10 K, is smaller than the CEF energy-level width (randomness) of $\sim$ 16 K. The pseudospin-$\frac{1}{2}$ magnetic moments can be calculated by,
\begin{equation}
m_\alpha^{i,j}=\mu_Bg_J\langle E_i|J_\alpha|E_j\rangle,
\label{eq13}
\end{equation}
under the subspace of $|E_1\rangle$ and $|E_2\rangle$,
\begin{equation}
m_x=m_y=\left(
       \begin{array}{cc}
         0 & 0 \\
         0 & 0 \\
       \end{array}
     \right),
\label{eq14}
\end{equation}
\begin{equation}
m_z=\mu_B\left(
       \begin{array}{cc}
         0 & 2.27 \\
         2.27 & 0 \\
       \end{array}
     \right).
\label{eq15}
\end{equation}
As a result, the final CEF ground-state quasi-doublet can be obtained,
\begin{equation}
|\sigma_{\pm}\rangle=\frac{1}{\sqrt{2}}(|E_1\rangle\pm|E_2\rangle),
\label{eq16}
\end{equation}
with the Ising moments along the \emph{c}-axis, $\pm$ 2.27$\mu_B$, respectively ($g_a^{\rm CEF}=g_b^{\rm CEF}=0$ and $g_c^{\rm CEF}\sim4.53$).

\bibliography{Pr_1}

\end{document}